# R-Index: A Robust Metric for IVIM Parameter Estimation on Clinical MRI Scanners

**Author Names and Degrees:**

Yan Dai, BS[1]; Xun Jia, PhD[2]; Yen-peng Liao, PhD[1]; Jie, Deng, PhD[1][†]

[†]Corresponding author

**Affiliations:**

[1] Department of Radiation Oncology, University of Texas Southwestern Medical Center, Dallas, TX, US

[2] Department of Radiation Oncology and Molecular Radiation Sciences, Johns Hopkins University, Baltimore, MD, USA

**Corresponding Author:**

Jie Deng, PhD

University of Texas Southwestern Medical Centre

2280 Inwood Rd, Dallas, TX, 75235, US

Telephone: (214) 645-5140

Email: Jie.Deng@UTSouthwestern.edu

**Acknowledgments:**

The authors would like to thank Xiaoyu Hu, PhD, for his valuable feedback on the manuscript, particularly his suggestions for improving the clarity and expression of the text. The authors also acknowledge the use of OpenAI's ChatGPT for assisting in refining the language.

**Grant Support:**

This work was supported in part by NIH/NCI R37CA214639, NIH/NIBIB R01EB032716, and NIH/NCI R01285379.

**Running Title:** *R*-Index for Clinical IVIM MRI

**Abstract:**

**Background:** Intravoxel Incoherent Motion (IVIM) model characterizes both water diffusion and perfusion in tissues, providing quantitative biomarkers valuable for tumor tissue characterization. However, parameter estimation based on this model is challenging due to its ill-posed nature, resulting in poor reproducibility, particularly at low signal to noise ratios (SNRs) in a clinic scenario.

**Purpose:** This study analyzes the uncertainty of IVIM model fitting, quantifies parameter collinearity, and introduces a new index with enhanced robustness to enhance clinical applicability of the IVIM model.

**Study Type:** Prospective.

**Population:** One healthy volunteer.

**Field Strength/Sequence:** 1.5T; single-shot EPI DWI.

**Assessment:** The probability distributions of estimated IVIM parameters were evaluated across a clinically relevant range. Collinearity among parameters was assessed and a new metric, the R-index, was proposed. The R-index linearly combines individual IVIM parameters to mitigate collinearity and reduce estimation uncertainty. Simulation and a volunteer study was conducted to validate the presence of parameter collinearity and to assess the robustness of the R-index.

**Statistical Tests:** N/A

**Results:** In simulation studies with a typical clinical setting (SNR = 20), normalized IVIM parameters exhibited mean standard deviations ranging from 0.107 to 0.269, while the R-index showed a reduced deviation of 0.064. Repeated scans in a healthy volunteer confirmed the presence of parameter collinearity, with 32% of voxels exhibiting statistically significant correlations ($p < 0.05$) among fitted IVIM parameters, and a mean Pearson correlation coefficient of $r = -0.96$.

**Data Conclusion:** The R-index provides a robust metric for IVIM model fitting under low SNR conditions typical of clinical MRI, offering improved reproducibility and potential for broader clinical applicability.

**Keywords:** Quantitative MRI; Intravoxel Incoherent Motion model; model fitting; robustness

## 1. Introduction

Diffusion-weighted imaging (DWI) is a quantitative MRI technique that probes microscopic tissue structure by measuring the incoherent motion of water molecules (1). The Intravoxel Incoherent Motion (IVIM) model (2), as shown in Equation 1, is a two-compartment model attributing the DWI signal decay to a slow component from water diffusion and a fast component from microcirculation:

$$\frac{S}{S_0} = f \cdot exp(-D_{\text{p}}b) + (1 - f) \cdot \exp(-D_{\text{t}}b). \qquad [1]$$

Here $S_0$ is the signal without diffusion weighting ($b = 0$), $f$ and $D_{\text{p}}$ represent the volume fraction and diffusivity of water moving within the capillary network, and $D_{\text{t}}$ denotes the true diffusivity of water in the extracellular space (3,4). Estimating $f$, $D_{\text{p}}$, and $D_{\text{t}}$ by fitting the IVIM model to DWI signals has been widely studied for applications such as tumor characterization (5-7), cerebral perfusion assessment (8,9), and treatment monitoring in oncology(10,11).

However, clinical translation of the IVIM model has been hindered by poor reproducibility (12-15). This challenge arises from various sources (16,17), with one of the key issues being uncertainty in parameter estimation, particularly under low signal-to-noise ratio (SNR) conditions (18-21), which are common due to scan time constraints and physiological motion (8,22,23). Several algorithms have been introduced to improve IVIM fitting stability. The segmented least squares method fits mono-exponential decays to high and low b-value ranges independently, requiring an empirically selected cutoff b-value that introduces subjective bias (24,25), particularly when $D_{\text{t}}$ and $D_{\text{p}}$ are similar (24,26). Bayesian approaches incorporate prior knowledge to stabilize the fit, though such priors may lack generalizability across tissue types (5). More recently, physics-informed neural networks such as IVIMNET (27) have been proposed to solve the inverse problem via unsupervised learning.

The difficulty in fitting the IVIM model lies fundamentally in the ill-conditioned nature of the bi-exponential model. It contains two non-orthogonal exponential components that are inherently difficult to disentangle, especially when $D_p$ and $D_t$ are similar (28,29). Most parameter estimation algorithms minimize some residual loss function, but in bi-exponential models, the loss function often shows weak sensitivity to the parameters, a phenomenon known as model "sloppiness" (30,31) . As a result, estimation becomes highly sensitive to noise, and fitted parameters can vary widely across repeated measurements (32,33). Several tools have been developed to analyze this "sloppiness" (34-37) . Monte Carlo simulation is a straightforward method to characterize the uncertainty by repeatedly simulating noisy measurements and performing data fitting. Bayesian inference can be used to generate posterior distributions and inter-parameter correlations(36) . The profile likelihood method perturbs one parameter at a time while holding others fixed to study local curvature of the cost function (38) , and asymptotic methods offer analytical approximations of local variability through likelihood function expansion (39) . Local geometric tools such as the Fisher Information Matrix quantify parameter uncertainty and collinearity.

In quantitative MRI, understanding the model manifold and the statistical behavior of fitted parameters is essential to avoid misleading biomarkers, particularly in IVIM modeling under clinically achievable SNRs (40). In this study, we identify a strong and consistent collinearity between $f$ and $D_t$, which limits their independent estimation. To address this, we propose the R-index, a linear combination of $f$ and $D_t$ that cancels their mutual collinearity and improves robustness under clinical conditions. Both theoretical analysis and experimental validation demonstrate that the R-index offers superior repeatability and serves as a stable surrogate metric. Our main contribution is the quantification of intrinsic parameter collinearity in IVIM and the introduction of the R-index as a robust imaging biomarker.

## 2. Materials and Methods

### 2.1 Probability distribution of estimated parameters

To analyze the parameter fitting of IVIM model, we studied the probability distribution of estimated parameters $\boldsymbol{\theta}_{\text{est}}$ for a given set of ground truth IVIM parameters $\boldsymbol{\theta}_{\text{gt}}$. As is expresses as Equation 2

$$P(\boldsymbol{\theta}_{\text{est}}|\boldsymbol{\theta}_{\text{gt}}) = \int_{\boldsymbol{y}_{\text{m}}} P(\boldsymbol{\theta}_{\text{est}}|\boldsymbol{y}_{\text{m}}; \boldsymbol{\theta}_{\text{gt}})P(\boldsymbol{y}_{\text{m}}|\boldsymbol{\theta}_{\text{gt}}) \, \text{d}\boldsymbol{y}_{\text{m}}, \qquad [2]$$

where $\boldsymbol{\theta}_{\text{est}}$ and $\boldsymbol{\theta}_{\text{gt}}$ are vectors $(f, D_{\text{t}}, D_{\text{p}})$ for the IVIM model. $\boldsymbol{y}_{\text{m}}$ is a vector of measured DWI signal intensities $S_i$ normalized by the signal at $b = 0 \text{ mm}^2/\text{s}$, denoted as $(S_{b1}/S_0, S_{b2}/S_0, \ldots, S_{b_M}/S_0)$. Assuming the noise in $S_i$ follows a Rician distribution, which can be approximated as Gaussian distribution for SNR above 5.0 (41), the noise in the normalized measurements $\boldsymbol{y}_{\text{m}}$ will be the ratio of two Gaussian distributions. We further assumed the noise in $\boldsymbol{y}_{\text{m}}$ as a Gaussian distribution, given that the SNRs of $S_0$ and $S_b$ satisfy the condition that $\sigma_0^2/S_0^2 + \sigma_b^2/S_b^2 \leq 1$, where $\sigma_i$ is the standard deviation of the Gaussian noise for signal $S_i$ (42). Under this assumption, the probability of the measured signal $\boldsymbol{y}_{\text{m}}$ arising from the known ground truth IVIM parameter $\boldsymbol{\theta}_{\text{gt}}$ is given by:

$$P(\boldsymbol{y}_{\mathrm{m}}|\boldsymbol{\theta}_{\mathrm{gt}}) = \prod_{i=1}^{M} \mathcal{N}\left(y_{\mathrm{m},i}; y_{\mathrm{gt},i}, \sigma_i\right),$$ [3]

where $\mathcal{N}(.)$ represents the Gaussian distribution with a mean of $\boldsymbol{y}_{\mathrm{gt}}$, the noiseless normalized IVIM signal based on parameters $\boldsymbol{\theta}_{\mathrm{gt}}$ computed using Equation 1.

The term $P(\boldsymbol{\theta}_{\mathrm{est}}|\boldsymbol{y}_{\mathrm{m}})$ is the distribution of the estimated parameter $\boldsymbol{\theta}_{\mathrm{est}}$ given the measurement $\boldsymbol{y}_{\mathrm{m}}$, which depends on specific data fitting algorithms. In this study, we estimated it using the Bayes equation:

$$P(\boldsymbol{\theta}_{\mathrm{est}}|\boldsymbol{y}_{\mathrm{m}}) = \frac{P(\boldsymbol{y}_{\mathrm{m}}|\boldsymbol{\theta}_{\mathrm{est}})P(\boldsymbol{\theta}_{\mathrm{est}})}{P(\boldsymbol{y}_{\mathrm{m}})} \propto P(\boldsymbol{y}_{\mathrm{m}}|\boldsymbol{\theta}_{\mathrm{est}})P(\boldsymbol{\theta}_{\mathrm{est}}).$$ [4]

Assuming a uniform prior distribution for $P(\boldsymbol{\theta}_{\mathrm{est}})$, the probability distribution $P(\boldsymbol{\theta}_{\mathrm{est}}|\boldsymbol{y}_{\mathrm{m}})$ is proportional to $P(\boldsymbol{y}_{\mathrm{m}}|\boldsymbol{\theta}_{\mathrm{est}})$, following a multi-component Gaussian distribution:

$$P(\boldsymbol{\theta}_{\mathrm{est}}|\boldsymbol{y}_{\mathrm{m}}) \propto P(\boldsymbol{y}_{\mathrm{m}}|\boldsymbol{\theta}_{\mathrm{est}}) = \prod_{i=1}^{M} \mathcal{N}\left(y_{\mathrm{m},i}; y_{\mathrm{est},i}, \sigma_i\right),$$ [5]

. Here $\boldsymbol{y}_{\mathrm{est}}$ is the IVIM signal based on parameters $\boldsymbol{\theta}_{\mathrm{est}}$ computed using Equation 1.

In this study, we considered DWI data acquisition with nonzero b-values of $5, 50, 100, 200, 450, 800, 1000 \, s/\mathrm{mm}^2$. The standard deviations of noise at all b-values were set to 0.05, given $S_0 = 1$, representing a typical noise level with SNR~20. We analyzed the ground truth IVIM parameter $\boldsymbol{\theta}_{\mathrm{gt}}$ within the range of $f \in (0, 0.3), D_{\mathrm{t}} \in (0, 0.003) \, \mathrm{mm}^2/s, \text{and } D_{\mathrm{p}} \in (0.003, 0.05) \, \mathrm{mm}^2/s$ based on literatures(2,3). For each $\boldsymbol{\theta}_{\mathrm{gt}}$, we computed the distribution $P(\boldsymbol{\theta}_{\mathrm{est}}|\boldsymbol{\theta}_{\mathrm{gt}})$ numerically via Equation 2. Specifically, to compute the integral over the measured signal $\boldsymbol{y}_m$, we used a Monte Carlo-based numerical integration approach. One issue is the normalization factor in Equation 4 is missing. To mitigate this issue, given that $\int_{\boldsymbol{\theta}_{\mathrm{est}}} P(\boldsymbol{\theta}_{\mathrm{est}}|\boldsymbol{y}_{\mathrm{m}}) \, \mathrm{d}\boldsymbol{\theta}_{\mathrm{est}} = 1$, we computed $P(\boldsymbol{y}_{\mathbf{m}}|\boldsymbol{\theta}_{\mathrm{est}})$ according to Equation 5 over a $100 \times 100 \times 100$ grid of $\boldsymbol{\theta}_{\mathrm{est}}$ evenly distributed within the range of $f \in (0, 1.0), D_{\mathrm{t}} \in (0, 0.003) \, \mathrm{mm}^2/s, \text{and } D_{\mathrm{p}} \in (0.003, 0.05) \, \mathrm{mm}^2/s$, resembling a common IVIM parameter estimation fitting scenario (5,6,18,19,24,43). These values were then summed up to obtain the normalization factor.

## 2.2 Analysis on estimated parameter stability and the proposed R-index

Since the absolute variances of IVIM parameters are influenced by their scales and units, which do not directly reflect their utility as quantitative biomarkers, we normalize the IVIM parameters using a scaling vector $v = (1.0, 0.003 \, \mathrm{mm}^2/s, 0.05 \, \mathrm{mm}^2/s)$ for $(f, D_{\mathrm{t}}, D_{\mathrm{p}})$ before the calculation of covariance matrix. The normalized parameters are denoted as $\widehat{\boldsymbol{\theta}} = (\hat{f}, \widehat{D}_{\mathrm{t}}, \widehat{D}_{\mathrm{p}})$ as dimensionless parameters, bringing them to the range of $(0, 1.0)$.

The stability of estimated parameters was then analyzed. For each $\widehat{\boldsymbol{\theta}}_{\mathrm{gt}}$ we computed the corresponding probability distributions $P(\widehat{\boldsymbol{\theta}}_{\mathrm{est}}|\widehat{\boldsymbol{\theta}}_{\mathrm{gt}})$ following the previously described steps. Then the $3 \times 3$ covariance matrix was calculated based on the sampled $\widehat{\boldsymbol{\theta}}_{\mathrm{est}}$. The diagonal values of the matrix represent the variance of the normalized estimated parameters, whereas nonzero off-diagonal elements indicate the collinearity among them. Such collinearity between the estimated parameters can be cancelled through a linear combination of them. To identify the linear combination leading to a smallest variance, we performed a principle component decomposition of the covariance matrix at each $\widehat{\boldsymbol{\theta}}_{\mathrm{gt}}$. The eigenvector $\boldsymbol{u}_{\mathrm{min}}(\widehat{\boldsymbol{\theta}}_{\mathrm{gt}})$ with the smallest eigen value $\lambda_{\mathrm{min}}(\widehat{\boldsymbol{\theta}}_{\mathrm{gt}})$ defines the coefficients for linearly combining the estimated parameters that leads to the most stable metric cancelling collinearity between the original IVIM parameters, forming $\boldsymbol{u}_{\mathrm{min}}(\widehat{\boldsymbol{\theta}}_{\mathrm{gt}})^{T} \widehat{\boldsymbol{\theta}}_{\mathrm{est}}$ .

To analysis the collinearity relationship between estimated parameters $\widehat{\boldsymbol{\theta}}_{\mathrm{est}}$ thoroughly across the whole feasible parameter space, $8 \times 8 \times 8$ $\widehat{\boldsymbol{\theta}}_{\mathrm{gt}}$ were uniformly sampled within the predefined range. The dispersion of eigen vectors $\boldsymbol{u}_{\mathrm{min}}$ for different sampled $\widehat{\boldsymbol{\theta}}_{\mathrm{gt}}$ were analyzed to check the general collinearity between across the feasible parameter space. Mean resultant length was calculated as a numerical metric quantifying the dispersion of $\boldsymbol{u}_{\mathrm{min}}$ . Such operation was performed across different SNR levels to analysis the trend of parameter collinearity. It will be shown later that the $\boldsymbol{u}_{\mathrm{min}}$ being similar for different $\widehat{\boldsymbol{\theta}}_{\mathrm{gt}}$ and measurement SNR, and this is the rationale of proposing the general metric, R-(Robust) -index$= \overline{\boldsymbol{u}}^{T} \widehat{\boldsymbol{\theta}}_{\mathrm{est}}$, where $\overline{\boldsymbol{u}} = \langle \boldsymbol{u}_{\mathrm{min}}(\widehat{\boldsymbol{\theta}}_{\mathrm{gt}}) \rangle$ is the vector averaged over all eigen vector over $\boldsymbol{u}_{\mathrm{min}}(\widehat{\boldsymbol{\theta}}_{\mathrm{gt}})$. R-index was proposed as an applicable metric to cancel the collinearity between parameter estimations, leading to the robustly estimated based on IVIM model,

The variance of the normalized estimated IVIM parameters $\widehat{\boldsymbol{\theta}}_{\mathrm{est}}$ and the proposed metric $R$ across the $8 \times 8 \times 8$ grid of different $\widehat{\boldsymbol{\theta}}_{\mathrm{gt}}$s was also visualized and compared to evaluate the improvement of repeatability by cancelling the parameter collinearity.

### 2.3 Evaluation

### 2.3.1 Simulation study validating the proposed distribution $P(\widehat{\boldsymbol{\theta}}_{\mathrm{est}}|\widehat{\boldsymbol{\theta}}_{\mathrm{gt}})$

A simulation study was conducted to validate the accuracy of the computed probability distribution $P(\widehat{\boldsymbol{\theta}}_{\mathrm{est}}|\widehat{\boldsymbol{\theta}}_{\mathrm{gt}})$. Gaussian noise with a standard deviation of 0.05 to both the real and imaginary part

of the simulated noiseless signals based on IVIM model at a series of b-values $0, 5, 50, 100, 200, 450, 800, 1000\ s/\text{mm}^2$. Signals at nonzero b-values were then normalized by dividing them by the signal at $b = 0\ s/\text{mm}^2$. The IVIM parameters were then estimated using the Trust Region Reflective algorithm by minimizing the sum of squared residuals within predefined boundaries. The fitting ranges were set as $\hat{f} \in (0, 1.0)$, $\widehat{D}_t \in (0, 1.0)$, and $\widehat{D}_p \in (0.1, 1.0)$ (corresponding to $f \in (0, 1.0)$, $D_t \in (0, 0.003)\ \text{mm}^2/s$, and $D_p \in (0.003, 0.05)\ \text{mm}^2/s$). Estimations reaching the fitting boundaries were discarded as fitting failures. 50000 noisy signals were generated and fitted, and the probability distribution was then generated via kernel density estimation with a Gaussian kernel. For visual comparison we plotted the 95% highest density region (HDR) of the both the theoretically derived and the numerically simulated distributions based on for three different representative $\widehat{\theta}_{\text{gt}}$. The agreement between pairs of distributions was quantified through the normalized Jensen-Shannon divergence between them. Such evaluation was performed for three different SNR levels $(10, 20, 40)$.

### 2.3.2 Simulation study validating parameter collinearity

To validate the general applicability of the proposed R-index in its effectiveness for cancelling parameter collinearity and reducing measurement variance, the distribution $P(\widehat{\theta}_{\text{est}}|\widehat{\theta}_{\text{gt}})$ was generated for $\widehat{\theta}_{\text{gt}}$ sampled across the predefined range for different SNR levels (5, 10, 20, 30, 40, 50). The variance for normalized parameters and the variance for R-index was calculated. For a more intrinsic comparison, the expected variance of the linear combination of IVIM parameters assuming that they are independent from each other is also computed.

### 2.3.3 Human subject study

To further evaluate the validity of parameter collinearity and the proposed R-index in clinical application, we conducted an experiment on a healthy volunteer brain under the approval from our Institutional Review Board. The volunteer was scanned repeatedly for 4 times on a clinical 1.5T MR simulator (MR-Sim) (Philips Healthcare, Amsterdam, Netherlands) used for radiation therapy simulation. The MR-sim scanner is a conventional diagnostic MRI scanner with a helium-free BlueSeal magnet and high-performance gradient (maximum gradient strengths of 40-45 mT/m and slew rates of 200 T/m/s). The DWI was repeatedly acquired four times using a single-shot EPI sequence with identical acquisition parameters: FOV = 300×256 mm$^2$, pixel size = 2×2 mm$^2$, slice thickness = 4 mm, TR = 3900 ms, TE = 95 ms, EPI factor =55, and SENSE factor =

2.5. Eight b-values $(0, 5, 50, 100, 200, 450, 800, 1000 \ s/mm^2)$ were used, with a total acquisition time of 3 minutes 7 seconds, achieving mean SNR of 18 on MR-Sim.

For each scan, voxel-wise IVIM parameter maps $(f, D_t, D_p$ and $R)$ were estimated using the trust-region reflective algorithm, with fitting boundaries identical to those used in our previous simulation study. To remove the impact from pixels exhibits mono-exponential decay signal patters, which deviates from the IVIM model, Akaike information criterion (AIC) was applied for evaluation of signal fitting(44). The pixels where signals fit better to the mono-exponential model were described. The IVIM parameters were then estimated separately on the rest pixels for each of the four DWI acquisitions. Additionally, voxels in which any of the four repeated fits reached a parameter boundary were discarded to eliminate potential fitting failures.

For each remaining voxel, we assessed the collinearity between $\hat{f}$ and $\widehat{D}_t$ by calculating the Pearson correlation coefficient and corresponding p-value across the four acquisitions. To further evaluate the stability of the R-index compared to conventional IVIM parameters, we computed the standard deviation of each normalized parameter across the four scans, as well as the standard deviation of the R-index.

## 3　Results

### 3.1 Probability distributions of the estimated parameters $P(\hat{\theta}_{est}|\hat{\theta}_{gt})$ and simulation validation

Figure 1a-c illustrates the 95% HDR of the probability distribution $P(\hat{\theta}_{est}|\hat{\theta}_{gt})$, for three representative $\theta_{gt}$ at SNR=*20*, obtained through our theoretical derivation in Equation 2, while Figure 1d-f represents the corresponding distributions generated via Monte Carlo simulation study. A visual comparison between them reveals a close agreement in the shape and extent of the HDR. To quantitively assess this similarity, we computed the normalized Jensen-Shannon divergence between the two distributions, a metric with a value of 0 indicating identical distributions and 1 representing that the distributions have no overlap in probability mass, indicating complete separation. The normalized Jensen-Shannon divergence (JSD) was 0.09, 0.07, 0.07 respectively, supporting the validity of our theoretical model. Results for another 2 noise levels ($SNR = 10$ and 20) were shown in the supplementary as Figure S1 and Figure S2. In both cases, the JSD achieved remained below 0.15, demonstrating the effectiveness of the derived posterior distributions in capturing the fitting behavior across different SNRs and ground truth values.

As is shown in Figure 1a-c, the posterior distributions of IVIM parameters exhibit strong collinearity, particularly between $\hat{f}$ and $\hat{D}_t$. Meanwhile, $\hat{D}_p$ displays a broad distribution, which accounts for its notably low repeatability in clinic applications. While the $\boldsymbol{u}_{\min}$, shown as brown arrows, appear similar for different $\hat{\boldsymbol{\theta}}_{gt}$s.

### 3.2 IVIM parameter collinearity and the proposed R-index

We analyzed the distribution of $\boldsymbol{u}_{\min}$, the direction in IVIM parameter space along which the estimates show the narrowest dispersion. Figure 2a-f presents 2D histograms of the spherical coordinates derived from $\boldsymbol{u}_{\min}(\hat{\boldsymbol{\theta}}_{gt})$ across sampled ground-truth parameters $\hat{\boldsymbol{\theta}}_{gt}$ for each SNR (5, 10, 20, 30, 40, 50). The azimuthal angle was defined as $\Theta = arctan(\hat{D}_t/\hat{f})$ and the polar angle as $\Phi = arccos(\hat{D}_p)$, with all vectors normalized. A green dashed line marks $\Phi = \pi/2$, i.e., corresponding to alignment with the $\hat{f} - \hat{D}_t$ plane. At low SNRs, $\boldsymbol{u}_{\min}$ directions are more scattered. As SNR increases, the directions converge near $\Phi = \pi/2$, especially for SNR$\geq$ 10. This consistent alignment allows us to define a general R-index as the average direction $\bar{\boldsymbol{u}} = \langle \boldsymbol{u}_{\min}(\hat{\boldsymbol{\theta}}_{gt}) \rangle$, capturing the most stable direction in parameter space.

Table 1 shows the mean resultant length (MRL) together with the averaged vector $\bar{\boldsymbol{u}}$. All the MRL calculated approaches 1, indicates perfect alignment between the $\boldsymbol{u}_{\min}(\hat{\boldsymbol{\theta}}_{gt})$ with different $\hat{\boldsymbol{\theta}}_{gt}$. Since SNR around 20 is usually achieved in clinical applications, the R-index is defined as Equation 6. The $\hat{D}_p$ term was omitted due to its minimal contribution.

$$\boldsymbol{R} = \bar{\boldsymbol{u}}^T\hat{\boldsymbol{\theta}} = (0.644 * \hat{f} + 0.759 * \hat{D}_t + 0.096 * \hat{D}_p) \approx 0.644 * \hat{f} + 0.759 * \hat{D}_t. \qquad [6]$$

Using the previously estimated parameter distributions, we computed the standard deviations of normalized parameter $\hat{\boldsymbol{\theta}}_{est}$ and of the R-index across all sampled $\hat{\boldsymbol{\theta}}_{gt}$s at each SNR. Figure 3 shows the distribution of these standard deviations. For comparison, we included the expected mean standard deviation (Std) of the R-index assuming no correlation between $\hat{f}$ and $\hat{D}_t$, calculated as $\sqrt{\left(0.644 \times \text{mean Std of } \hat{f}\right)^2 + \left(0.759 \times \text{mean Std of } \hat{D}_t\right)^2}$ (orange dashed lines). Mean Std values are summarized in Table 2. The large variability in individual IVIM parameters underscores their uncertainty, while the reduced variability of the R-index demonstrates its robustness by canceling collinearity.

### 3.3 Validation of parameter collinearity in repeated volunteer scans

Repeated acquisitions in the volunteer study further confirm the collinearity between $\hat{f}$ and $\widehat{D}_{\mathrm{t}}$. Figure 4a shows five representative voxels with significant correlations (p $\leq$ 0.05). For each selected pixel, four fitted values of $\hat{f}$ and $\widehat{D}_{\mathrm{t}}$ from each acquisitions are plotted and connected in a two-dimensional scatter plot to demonstrate the collinearity and variability. This subplot highlights the consistency in correlation patterns across pixels and provides a more intuitive understanding of how measurement noise affect parameter stability. Figure 4b presents the voxel-wise Pearson correlation between $\hat{f}$ and $\widehat{D}_{\mathrm{t}}$, showing a strong overall negative correlation (mean $r = -0.96$), with 32% of voxels reaching statistical significance. This is consistent with the collinearity pattern observed in simulation. To assess robustness, Figure 4c compares the standard deviation of normalized IVIM parameters and the R-index across scans. The R-index shows substantially lower variability, underscoring its stability as a clinically relevant summary metric.

## 4  Discussion

This study examined the probability distributions of IVIM parameter estimates under noise to better understand the statistical behavior of this ill-conditioned fitting problem, which is an essential step for using IVIM in quantitative imaging biomarkers. We identified two key findings: (1) strong, persistent collinearity among IVIM parameters leads to broad distributions; and (2) this collinearity remains consistent across various ground truth values and SNRs. Based on this, we proposed the R-index, a linear combination of IVIM parameters that cancels collinearity— demonstrating improved stability under clinically realistic SNRs.

IVIM parameter fitting is fundamentally an inverse problem complicated by weak signal sensitivity to parameter changes. Ideally, the model should provide a one-to-one mapping between signal and parameters, but in IVIM, measurement noise can render distinct parameter sets indistinguishable, leading to degeneracy and unreliable estimates. To further investigate this issue, we generated a set of noiseless IVIM signals $S/S_0$ on seven b-values (5,50,100,200,450,800,1000) s/mm$^2$ across a range of groud truth parameters $(f, D_{\mathrm{t}}, D_{\mathrm{p}})$, forming a signal matrix $S \in R^{7 \times N}$, with each column corresponding to an IVIM signal at one model parameter, and $N$ is the total number of signals considered. We performed principal component analysis (PCA) of this matrix, yielding $S = \bar{S} + U\Sigma V^T$, where $\bar{S}$ is the mean signal, $U$ is an unitary matrix with columns $u_i$ being the eigen vectors as the basis to represent signal variation, $\Sigma$ is diagnal matrix with entries $v_i$, and $V$ is a matrix with each column $v(\boldsymbol{\theta})$ containing coefficients for linearly representing the signal at $\boldsymbol{\theta}$ with collumns in $U$. These signals, i.e. columns of $S$ were very similar, as the mean signal accounted for 98% of the total signal intenstiy, defined as $|\bar{S}|/\sqrt{|S|_F^2/N}$, where $|.|$ is the vector 2-norm and $|.|_F$ is the matrix Frobenius norm. The remaining 2% signal variation contained useful information to derive the IVIM model parameters. Among the remaining signal variation, the first principal component accounted for 93% of it, with an amplitude of $v_1 \approx 0.117 \approx 0.342^2$, while the second and third components contributed only 6% ($v_2 \approx 0.008 \approx 0.089^2$) and 0.4% ($v_3 \approx 0.0005 \approx 0.022^2$), respectively. In the presence of noise with a typical SNR = 20, the amplitude of measurement noises is 0.05.Considering two signal with difference twice the noise amplityde being distinguishable, only the first principal component may be reliably measured.

The first two coefficients $v_i(\boldsymbol{\theta})$, $i = 1,2$ were plotted in Figure 5a,b. It is interesting to observe that the function $v_1(\boldsymbol{\theta})$ is approximately of a linear form. Fitted $v_i$ with a linear function of $\boldsymbol{\theta}$, yields Equation 7:

$$v_1(\boldsymbol{\theta}) = 1.019\hat{f} + 1.227\widehat{D}_{\text{t}} + 0.085\widehat{D}_{\text{p}} - 0.808 = 1.582\big(0.644\hat{f} + 0.776\widehat{D}_{\text{t}} + \qquad [7]$$
$$0.054\widehat{D}_{\text{p}}\big) - 0.808 \,,$$

with a residual of $R^2 = 0.008$. Ignoring the constant term and the scaling factor, this function closely matched the previously defined R-index in Equation 6. To better visualize the correlation between R-index and $v_1$, Figure 5c plots $v_1(\boldsymbol{\theta})$ against the R-index across all sampled $\boldsymbol{\theta}_{\text{gt}}$s. A strong Pearson correlation of r=0.97 with p~0 was observed. These findings suggest that under clinical SNR, the IVIM model effectively reduces to a system with a single resolvable degree of freedom, and the R-index aligns with this dominant axis of variation. Thus, the R-index serves as a robust and representative metric for summarizing IVIM signal behavior in noise-limited conditions.

Such correlation between $v_1(\boldsymbol{\theta})$ and R-index is further explained below. Consider the signal equation $S = S(\boldsymbol{\theta})$. Let $\boldsymbol{J} = \partial S/\partial \boldsymbol{\theta}$ be the Jacobian matrix. For a noise signal $\delta S = \epsilon$, the parameter estimation is perturbed that $\epsilon = \boldsymbol{J}\delta\theta$. The covariance of the perturbation to the estimated parameter $\delta\theta$ is given by $\text{cov}(\delta\boldsymbol{\theta}) = E[\delta\boldsymbol{\theta}\delta\boldsymbol{\theta}^T] = E[\boldsymbol{J}^+\boldsymbol{\epsilon}(\boldsymbol{J}^+\boldsymbol{\epsilon})^T] = \boldsymbol{J}^+ cov(\boldsymbol{\epsilon})(\boldsymbol{J}^+)^T = \sigma^2\boldsymbol{J}^+(\boldsymbol{J}^+)^T$, where $\boldsymbol{J}^+$ is the Moore-Penrose pseudoinverse of $\boldsymbol{J}$. The direction for robust estimation of $\delta\boldsymbol{\theta}$ in the $\boldsymbol{\theta}$ space corresponds to the eigenvector associated with the smallest eigen value of $cov(\delta\boldsymbol{\theta})$, which is also the eigenvector with the largest eigen value for matrix $(\boldsymbol{J}^+(\boldsymbol{J}^+)^T)^{-1} = \boldsymbol{J}\boldsymbol{J}^T$. Based on PCA $\boldsymbol{S} = \bar{S} + U\Sigma V^T$, this direction is given by $\nabla v_1(\boldsymbol{\theta})$. Note that, in general, this direction depends on $\boldsymbol{\theta}$. Yet from our empirical observation that $v_1(\boldsymbol{\theta}) \approx w^T\boldsymbol{\theta}$, it followed that the direction $w$ in Eq. 7 defines this direction, which led to the agreement between this expression and the previously defined R-index in Eq. 6. Note that the approximately linear form of $v_1(\boldsymbol{\theta})$ plays a crucial role in making it feasible to define the R-index. Otherwise $\nabla v_1(\boldsymbol{\theta})$ would be $\boldsymbol{\theta}$ dependent, and it is not possible to define a globally valid index.

It is also worthwhile to point out that for the second principal component, shown in Figure 5b, appears to follow a nonlinear function $af * D_p + b$, where $a$ and $b$ are function parameters. When the noise in the DWI measurement can be reduced to a level to allow the resolution of this principal component, the product $f * D_p$ can be robustly estimated. This fact may explain the relevance of $f * D_p$ reported in previous work(9,11,45).

An important clarification is that the R-index was derived from purely robust consideration for the non-linear IVIM signal fitting problem, and it does not necessarily imply a direct biological meaning. It serves as be a stable perfusion weighted metric derived from the IVIM model, particularly under low SNR conditions where the estimated perfusion fraction lacks repeatability. While the perfusion fraction may be more interpretable at high SNR, its sensitivity to noise limits its clinical utility. By combining perfusion fraction and diffusion coefficient, the R-index reduces collinearity and stabilizes estimation. Conceptually, it can be seen as a coarse-grained parameter that, while not isolating perfusion, retains meaningful perfusion sensitivity and offers greater robustness in clinical settings.

Establishing its biological relevance requires further studies to correlate this metric with physiologically meaningful information. One encouraging fact is that the R-index proposed in our study closely resembles a hypoxia index reported in a previous study (43), where a metric $f/0.43 + D_t/(0.79 \times 10^{-3}) \approx 3.61 * (0.644\hat{f} + 1.050\widehat{D}_t)$ was used to assess the hypoxia levels in prostate cancer. This similarity suggests a potential application of the R-index as both a robust and biologically relevant parameter for hypoxia assessment.

The proposed R-index depends on specific acquisition settings, including the number and range of b-values and noise levels, which influence IVIM fitting performance.(28,36). Our uncertainty analysis and PCA were conducted under fixed experimental conditions; variations in acquisition may alter the R-index formulation (e.g., combination coefficients). While its conceptual validity remains. This study also assumes Gaussian noise with constant variance in normalized signals $S_b/S_0$, which may not hold at high b-values due to low SNR (42). Future work should explore protocol optimization to improve both parameter robustness and the biological relevance of the R-index.

## 5   CONCLUSION

We analyzed the probability distribution of IVIM parameter estimates and identified inherent collinearity among them. Based on this insight, we proposed the R-index, a new metric that mitigates collinearity and improves robustness under clinically achievable SNRs. Simulations and experimental results validated its superior stability compared to individual IVIM parameters.

**TABLE 1** Coefficients of the averaged vector $\bar{\boldsymbol{u}}$ and mean resultant length (MRL) across different SNRs

| SNR | Coefficient of $\hat{f}$ | Coefficient of $\widehat{D}_t$ | Coefficient of $\widehat{D}_p$ | MRL |
| --- | --- | --- | --- | --- |
| 5 | 0.815 | 0.518 | 0.261 | 0.98 |
| 10 | 0.714 | 0.679 | 0.169 | 0.98 |
| 20 | 0.644 | 0.759 | 0.096 | 0.99 |
| 30 | 0.621 | 0.781 | 0.066 | 0.99 |
| 40 | 0.615 | 0.787 | 0.050 | 0.98 |
| 50 | 0.614 | 0.788 | 0.041 | 0.98 |

The coefficients represent the contributions of each parameter dimension to the direction of minimal dispersion in the estimated parameter space. The direction tends to be consistent for SNR ≥ 20, with the coefficient of $\widehat{D}_p$ approaches 0. The MRL all approaches 1, indicating the narrow distribution of $\boldsymbol{u}_{\min}(\hat{\boldsymbol{\theta}}_{\mathrm{gt}})$ for all sampled $\hat{\boldsymbol{\theta}}_{\mathrm{gt}}$s.

**TABLE 2** Mean standard divation (Std) of IVIM parameters and R-index at different SNRs.

| SNR | Mean Std of $\hat{f}$ | Mean Std of $\widehat{D}_t$ | Mean Std of $\widehat{D}_p$ | Mean Std of $R$-index | Expected mean Std of $R$-index |
|---|---|---|---|---|---|
| 5 | 0.190 | 0.205 | 0.277 | 0.166 | 0.198 |
| 10 | 0.161 | 0.150 | 0.276 | 0.104 | 0.154 |
| 20 | 0.136 | 0.107 | 0.269 | 0.064 | 0.120 |
| 30 | 0.117 | 0.087 | 0.260 | 0.048 | 0.100 |
| 40 | 0.102 | 0.073 | 0.249 | 0.039 | 0.086 |
| 50 | 0.090 | 0.063 | 0.238 | 0.034 | 0.075 |

The mean Std of R-index has smaller than both the IVIM parameters and the expected mean Std when $\hat{f}$ and $\widehat{D}_t$ are independent across all SNRs, indicating the collinearity between IVIM parameters and the robust of R-index achieved by cancelling such collinearity.

**FIGURE 1 (a-c)** Probability distribution $P\left(\widehat{\boldsymbol{\theta}}_{\text{est}}\middle|\boldsymbol{\theta}_{\text{gt}}\right)$ generated based on Equation 2 for SNR=20 with $\boldsymbol{\theta}_{\text{gt}}$s marked as blue dots ( $\hat{f} = 0.05, \widehat{D}_{\text{t}} = 0.27, \widehat{D}_{\text{p}} = 0.3; \hat{f} = 0.15, \widehat{D}_{\text{t}} = 0.4, \widehat{D}_{\text{p}} = 0.5; \hat{f} = 0.25, \widehat{D}_{\text{t}} = 0.33, \widehat{D}_{\text{p}} = 0.7$, corresponding to $f = 0.05, D_{\text{t}} = 0.0008 \text{ mm}^2/\text{s}, D_{\text{p}} = 0.015 \text{ mm}^2/\text{s}; f = 0.15, D_{\text{t}} = 0.0012 \text{ mm}^2/\text{s}, D_{\text{p}} = 0.025 \text{ mm}^2/\text{s}; f = 0.25, D_{\text{t}} = 0.0010 \text{ mm}^2/\text{s}, D_{\text{p}} = 0.035 \text{ mm}^2/\text{s}$ respectively). Surfaces represents 95% highest density regions (HDR). Brown arrows indicate the directions of the eigenvectors $\boldsymbol{u}_{\text{min}}$. **(d-f)** Corresponding probability distributions from Monte Carlo simulations, with 95% HDR shown.

**FIGURE 2** Angular distributions of directions indicating minimal dispersion in IVIM parameter estimates at different SNR levels. Subplots **(a–f)** correspond to SNRs of 5, 10, 20, 30, 40, and 50, respectively. Each vector direction reflects the axis along which parameter estimates are least variable under noise. The spherical angles are defined as: $\Theta = arctan(\widehat{D}_{\text{t}}/\hat{f})$, $\Phi = arccos(\widehat{D}_{\text{p}})$

**FIGURE 3 (a–f)** Standard deviation (Std) distributions of IVIM parameters and the R-index at SNRs of 5, 10, 20, 30, 40, and 50, respectively. The expected Std of the R-index assuming independence between $\hat{f}$ and $\widehat{D}_{\text{t}}$ is shown as dashed orange lines. $\widehat{D}_{\text{p}}$ shows consistently high variability. The R-index demonstrates reduced Std, highlighting its robustness against parameter collinearity.

**FIGURE 4 (a)** Fitting results from four repeated scans in five representative voxels with significant correlations (p≤ 0.05). Each voxel's estimates are color-coded and connected with dashed lines. The joint variability of $\hat{f}$ and $\widehat{D}_{\text{t}}$ reflects their instability due to noise. **(b)** Density plot of Pearson correlation coefficient r vs. p-value between $\hat{f}$ and $\widehat{D}_{\text{t}}$ across all voxels. A strong negative correlation is observed in most cases. **(c)** Boxplots showing the Std across scans for normalized IVIM parameters and the R-index. The expected Std of the R-index under parameter independence is indicated by a dashed orange line. The R-index shows substantially lower variability, confirming its robustness.

**FIGURE 5** Spatial distribution of the first **(a)** and second **(b)** principal components of simulated IVIM signals sampled across the $\boldsymbol{\theta}_{\text{gt}}$ grid. **(c)** Scatter plot showing a strong linear correlation between the R-index and the first principal component across all sampled $\boldsymbol{\theta}_{\text{gt}}$s.

**a** 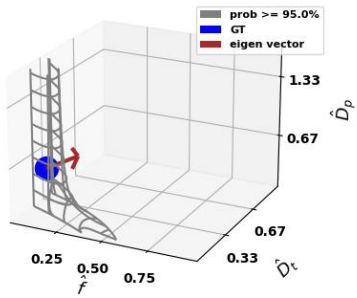

**b** 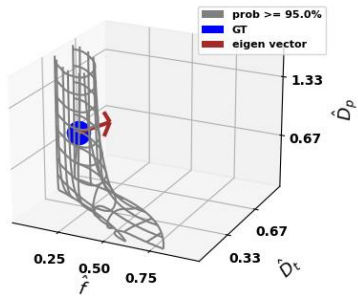

**c** 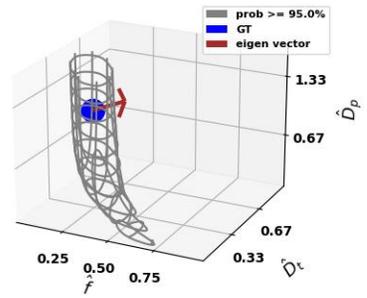

**d** 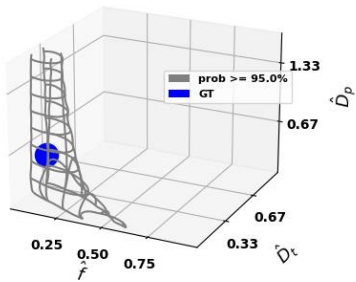

**e** 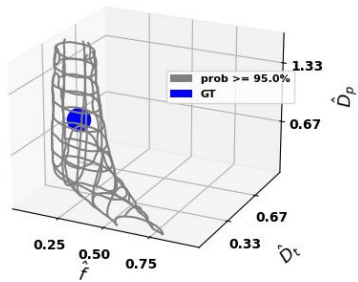

**f** 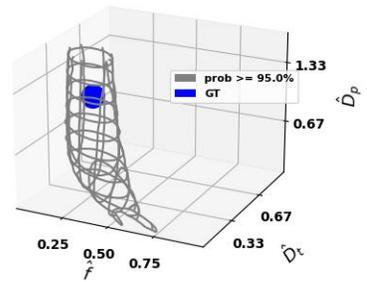

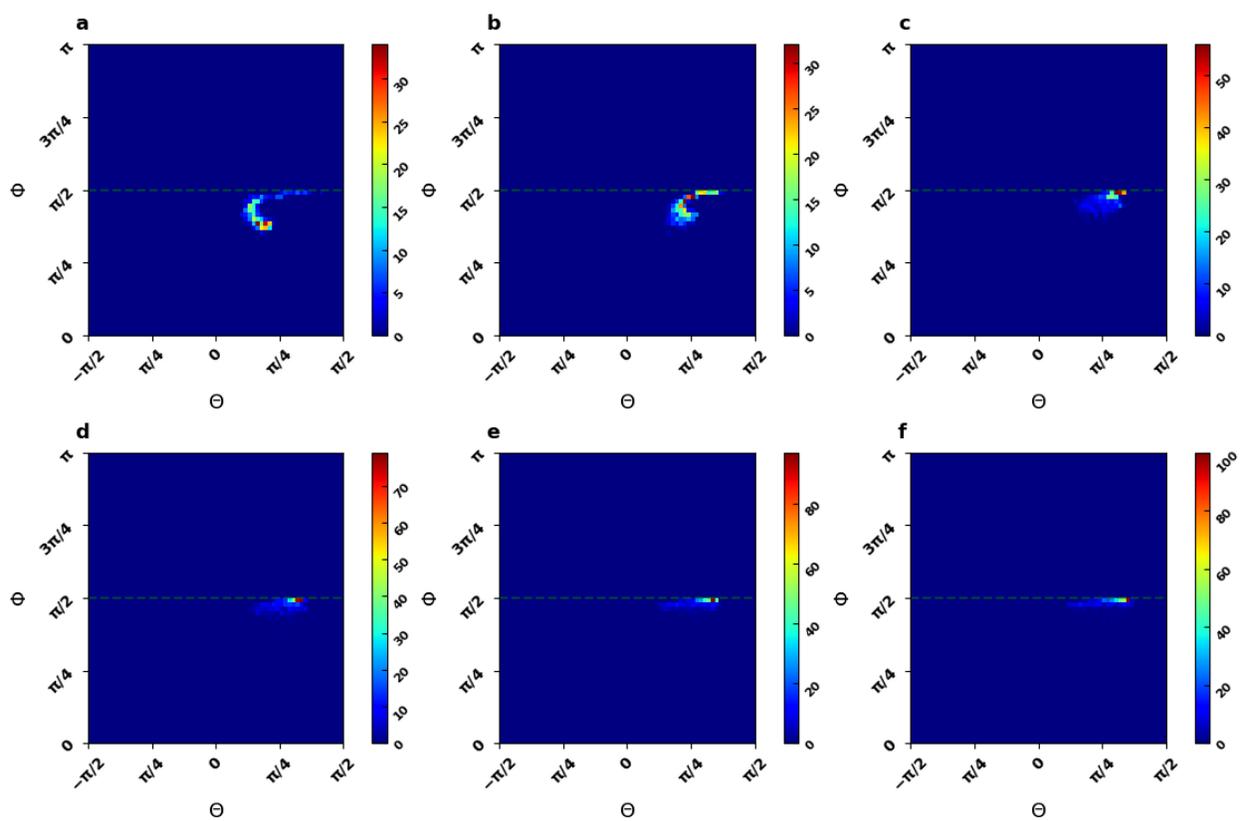

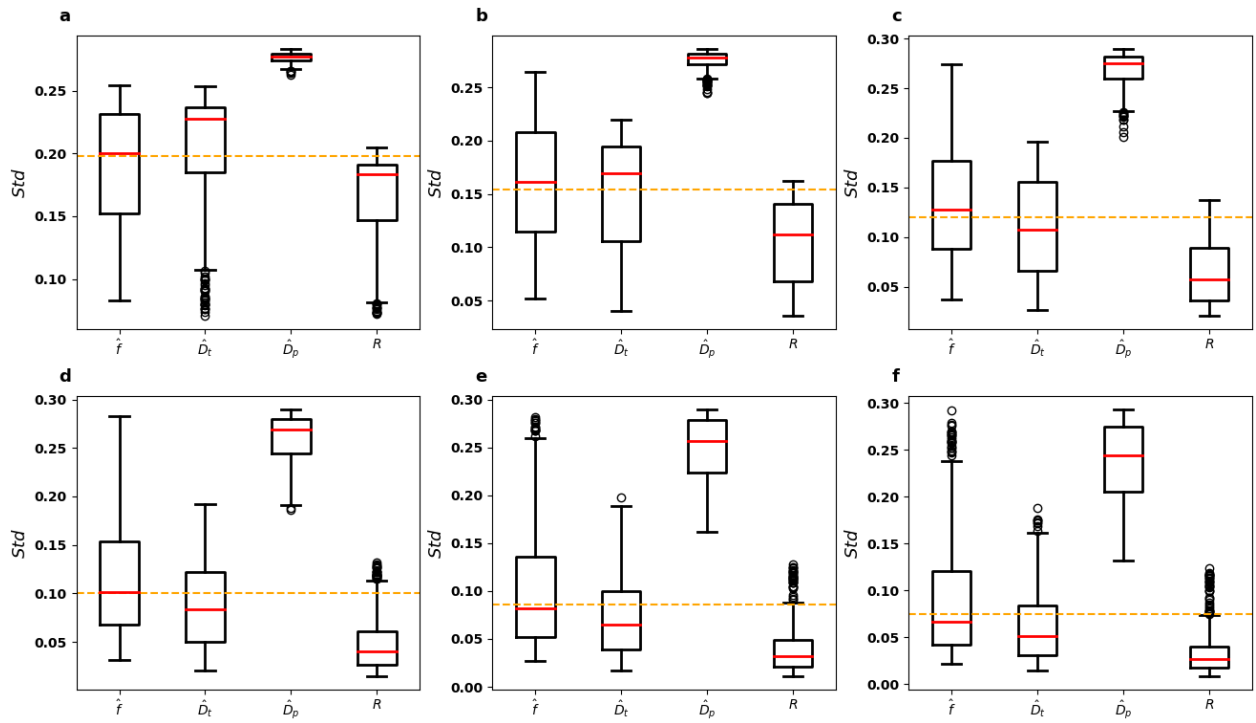

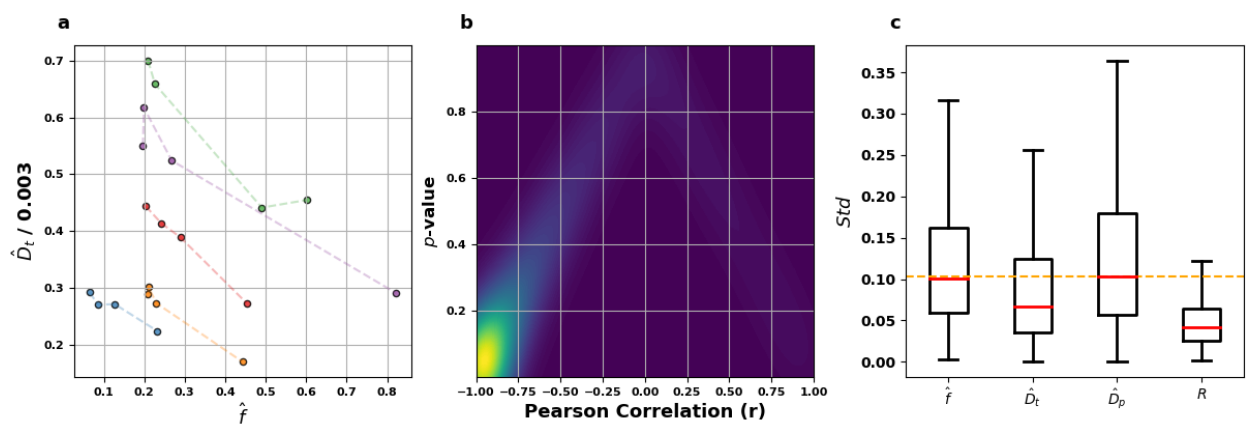

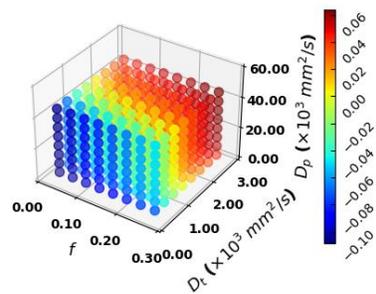
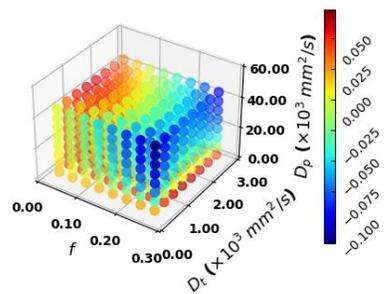
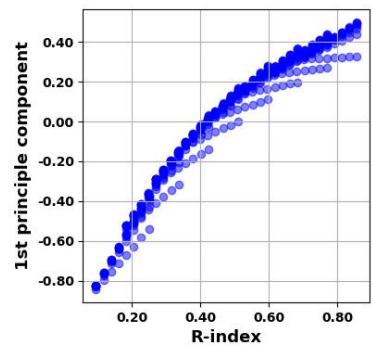

**Supplementary**

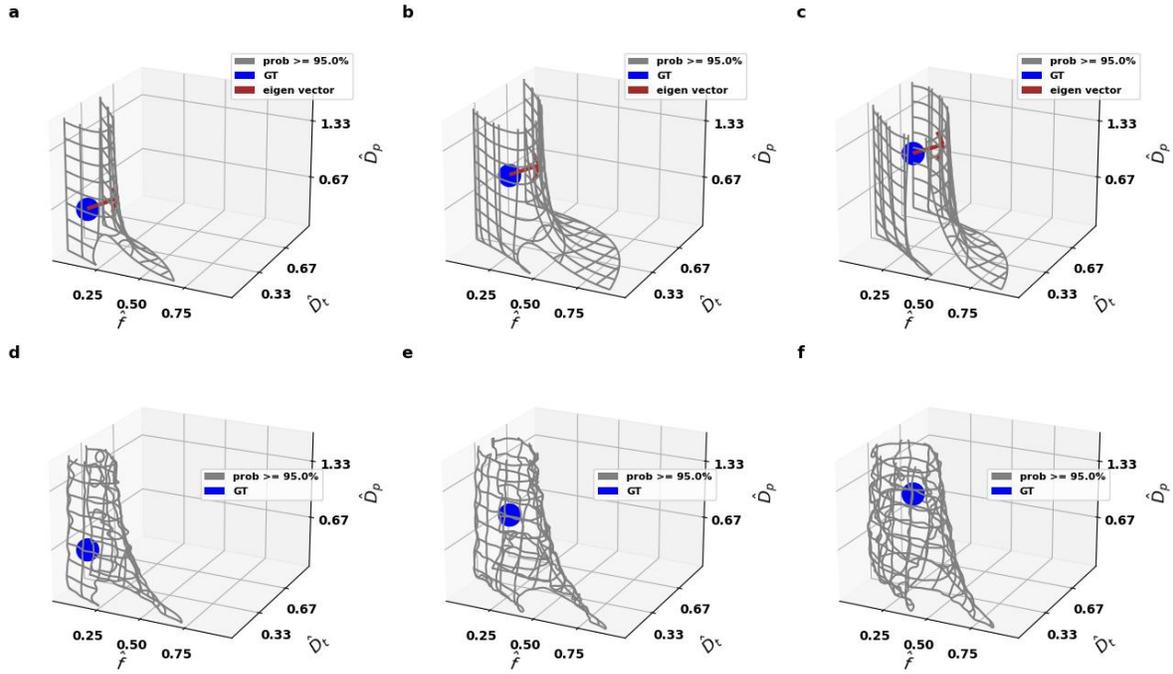

**FIGURE S1 (a-c)** Probability distribution $P\left(\widehat{\boldsymbol{\theta}}_{\text{est}} \big| \widehat{\boldsymbol{\theta}}_{\text{gt}}\right)$ generated based on Equation 2 for SNR=10 with $\widehat{\boldsymbol{\theta}}_{\text{gt}}$s marked as blue dots ($\hat{f} = 0.05, \widehat{D}_{\text{t}} = 0.27, \widehat{D}_{\text{p}} = 0.3$; $\hat{f} = 0.15, \widehat{D}_{\text{t}} = 0.4, \widehat{D}_{\text{p}} = 0.5$; $\hat{f} = 0.25, \widehat{D}_{\text{t}} = 0.33, \widehat{D}_{\text{p}} = 0.7$, corresponding to $f = 0.05, D_{\text{t}} = 0.0008$ mm²/s, $D_{\text{p}} = 0.015$ mm²/s; $f = 0.15, D_{\text{t}} = 0.0012$ mm²/s, $D_{\text{p}} = 0.025$ mm²/s; $f = 0.25, D_{\text{t}} = 0.0010$ mm²/s, $D_{\text{p}} = 0.035$ mm²/s respectively). Surfaces represents 95% highest density regions (HDR). Brown arrows indicate the directions of the eigenvectors $\boldsymbol{u}_{\text{min}}$. **(d-f)** Corresponding probability distributions from Monte Carlo simulations, with 95% HDR shown.

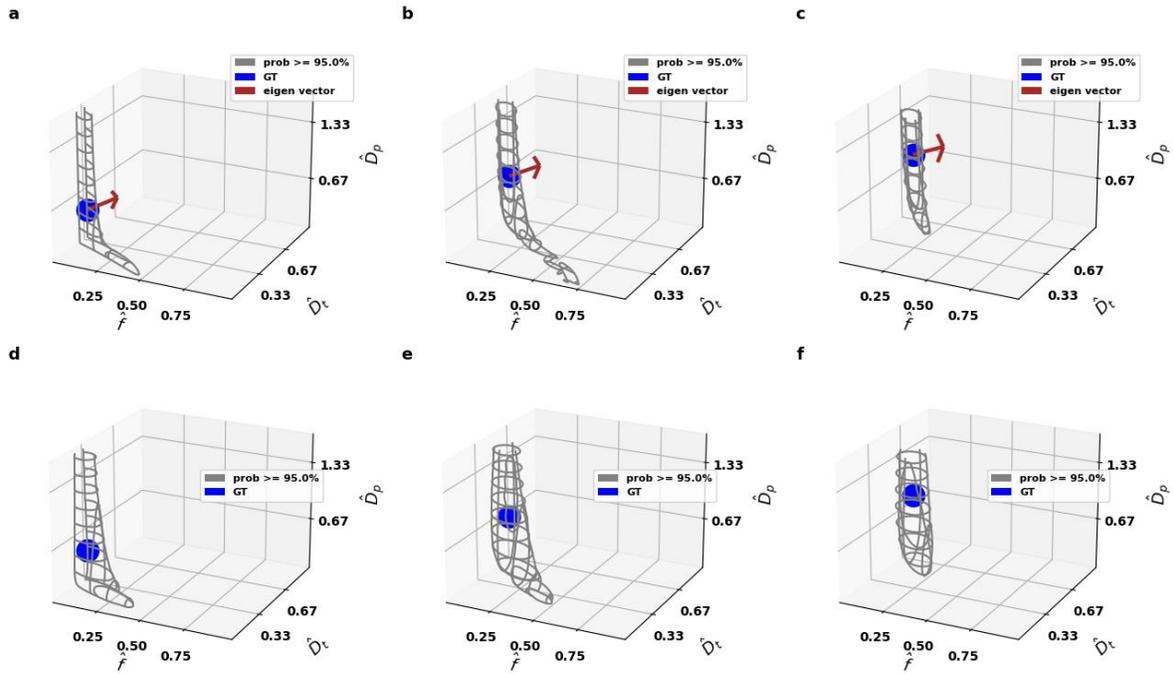

**FIGURE S2 (a-c)** Probability distribution $P\left(\widehat{\boldsymbol{\theta}}_{\text{est}} \middle| \widehat{\boldsymbol{\theta}}_{\text{gt}}\right)$ generated based on Equation 2 for SNR=40 with $\widehat{\boldsymbol{\theta}}_{\text{gt}}$s marked as blue dots ( $\hat{f} = 0.05, \widehat{D}_{\text{t}} = 0.27, \widehat{D}_{\text{p}} = 0.3$; $\hat{f} = 0.15, \widehat{D}_{\text{t}} = 0.4, \widehat{D}_{\text{p}} = 0.5$; $\hat{f} = 0.25, \widehat{D}_{\text{t}} = 0.33, \widehat{D}_{\text{p}} = 0.7$, corresponding to $f = 0.05, D_{\text{t}} = 0.0008 \text{ mm}^2/s, D_{\text{p}} = 0.015 \text{ mm}^2/s$; $f = 0.15, D_{\text{t}} = 0.0012 \text{ mm}^2/s, D_{\text{p}} = 0.025 \text{ mm}^2/s$; $f = 0.25, D_{\text{t}} = 0.0010 \text{ mm}^2/s, D_{\text{p}} = 0.035 \text{ mm}^2/s$ respectively). Surfaces represents 95% highest density regions (HDR). Brown arrows indicate the directions of the eigenvectors $\boldsymbol{u}_{\text{min}}$. **(d-f)** Corresponding probability distributions from Monte Carlo simulations, with 95% HDR shown.